\documentclass{article}
\usepackage{spconf,amsmath}
\usepackage{amssymb}

\usepackage{xurl}        % must come before hyperref
\usepackage{hyperref}

\usepackage{url}
\usepackage{enumitem} % It's fine to compress itemized lists if you used them in the manuscript

\setlist{nosep, leftmargin=14pt}

\usepackage{mwe} % to get dummy images
\usepackage{float} % Add this in your preamble
\usepackage{float} % Add this in your preamble
\usepackage{booktabs} % For \toprule, \midrule, etc.
\usepackage{threeparttable}
\usepackage{threeparttable} % For better table designs

\usepackage{graphicx} % for images
\title{AdaLoRA-QAT: Adaptive Low Rank and Quantization Aware Segmentation}
\name{Prantik Deb$^{\star}$, Srimanth Dhondy$^{\star}$, N. Ramakrishna$^{\dagger}$, Anu Kapoor$^{\dagger}$, Raju S. Bapi$^{\star}$, Tapabrata Chakraborti$^{\ddagger}$}

\address{
$^{\star}$ International Institute of Information Technology (IIIT-H), Hyderabad, India\\
$^{\dagger}$ Nizam’s Institute of Medical Sciences (NIMS), Hyderabad, India\\
$^{\ddagger}$ The Alan Turing Institute, London and University College London, United Kingdom}
\begin{document}
\maketitle
%\ninept
\begin{abstract}
Chest X-ray (CXR) segmentation is an important step in computer-aided diagnosis, yet deploying large foundation models in clinical settings remains challenging due to computational constraints. We propose AdaLoRA-QAT, a two-stage fine-tuning framework that combines adaptive low-rank encoder adaptation with full quantization-aware training. Adaptive rank allocation improves parameter efficiency, while selective mixed-precision INT8 quantization preserves structural fidelity crucial for clinical reliability. Evaluated across large-scale CXR datasets, AdaLoRA-QAT achieves 95.6\% Dice, matching full-precision SAM decoder fine-tuning while reducing trainable parameters by 16.6$\times$ and yielding 2.24$\times$ model compression. A Wilcoxon signed-rank test confirms that quantization does not significantly degrade segmentation accuracy. These results demonstrate that AdaLoRA-QAT effectively balances accuracy, efficiency, and structural trustworthiness, enabling compact and deployable foundation models for medical image segmentation. Code and pretrained models are available at: \url{https://prantik-pdeb.github.io/adaloraqat.github.io/}
\end{abstract}


\section{Introduction}
\label{sec:intro}

Chest radiography (CXR) is a widely accessible and cost-effective modality for screening pulmonary diseases such as pneumonia, tuberculosis, and COVID-19~\cite{mittal2017lung}. Accurate lung field segmentation is fundamental for isolating pulmonary parenchyma, enhancing abnormality visibility, enabling quantitative analysis, and improving the reliability of computer-aided diagnosis (CAD) systems. While deep learning models such as nnU-Net~\cite{isensee2018nnu}, DeepLabV3+~\cite{chen2018encoder}, and SegFormer~\cite{xie2021segformer} have achieved strong performance but robust generalization remains challenging due to anatomical variability, pathological distortions, and imaging artifacts, limiting clinical deployability.

Given a chest X-ray $\mathbf{x} \in \mathbb{R}^{H\times W\times3}$ and bounding-box prompt $\mathbf{b}$, we aim to predict a binary lung mask $\mathbf{y} \in \{0,1\}^{H\times W}$ under strict constraints on trainable parameters, memory footprint, and inference efficiency. To this end, we adapt the Segment Anything Model (SAM)~\cite{kirillov2023segment} using Parameter-Efficient Fine-Tuning (PEFT), leveraging Adaptive Low-Rank Adaptation (AdaLoRA)~\cite{zhang2023adalora} to dynamically allocate rank capacity to task-relevant transformer layers. To further enable practical deployment, we integrate Quantization-Aware Training (QAT), achieving low-bit precision while preserving fine structural fidelity.

We propose \emph{AdaLoRA-QAT}, a two-stage framework where Stage~1 learns adaptive and orthogonal low-rank subspaces in full precision, pruning redundant components to identify an efficient task-specific parameter space, and Stage~2 performs full-model quantization-aware fine-tuning while freezing rank masks, allowing stable adaptation to quantized constraints. This progressive strategy yields a compact, memory-efficient SAM variant that maintains diagnostic accuracy and robustness for real-world clinical use.

\textbf{Our main contributions are summarized as follows:} 

1) We introduce \emph{AdaLoRA-QAT}, a unified two-stage framework that couples adaptive low-rank encoder tuning with full model quantization-aware fine-tuning.

2) We design a mixed-precision strategy that selectively quantizes encoder feed-forward layers, decoder, and prompt encoder to INT8, while retaining attention QKV projections and AdaLoRA parameters ($\mathbf{P}$, $\mathbf{Q}$, $\mathbf{\Lambda}$) in FP32 to prevent rank collapse.

3) AdaLoRA-QAT achieves state-of-the-art efficiency: 95.6\% Dice with 16.6$\times$ parameter reduction and 2.24$\times$ model compression, with no significant performance degradation as validated by the Wilcoxon test.

% \end{enumerate}
%---------------------------------------------------------------------------------
\section{Method}

\subsection{Datasets}
\label{sec:dataset}
Our study utilizes publicly available chest X-ray (CXR) datasets including JSRT~\cite{shiraishi2000development}, QaTa-COV19~\cite{degerli2022osegnet}, COVID-19 Radiography~\cite{rahman2021exploring,chowdhury2020can}, Chest X-Ray Pneumothorax~\cite{siim-acr-pneumothorax}, and COVID-QU-Ex~\cite{tahir2021covid}. Together, these sources comprise 64,590 CXRs spanning diverse thoracic pathologies, providing a clinically representative benchmark for evaluating robustness and diagnostic generalization.

%-----------------------------------------------------------
% \begin{figure}[htb]
%   \centering
%   \includesvg[width=\columnwidth, height=8.8 cm]{images/AdaLoRA+FQAT.svg}
%   \vspace{-1em}
%   \caption{Two-stage training pipeline of the proposed AdaLoRA-QAT framework.}
%   \label{fig:pipeline_model}
% \end{figure}
% \vspace{-1.5em}
\begin{figure}[htb]
  \centering
  \includegraphics[width=\columnwidth, height=8.8 cm]{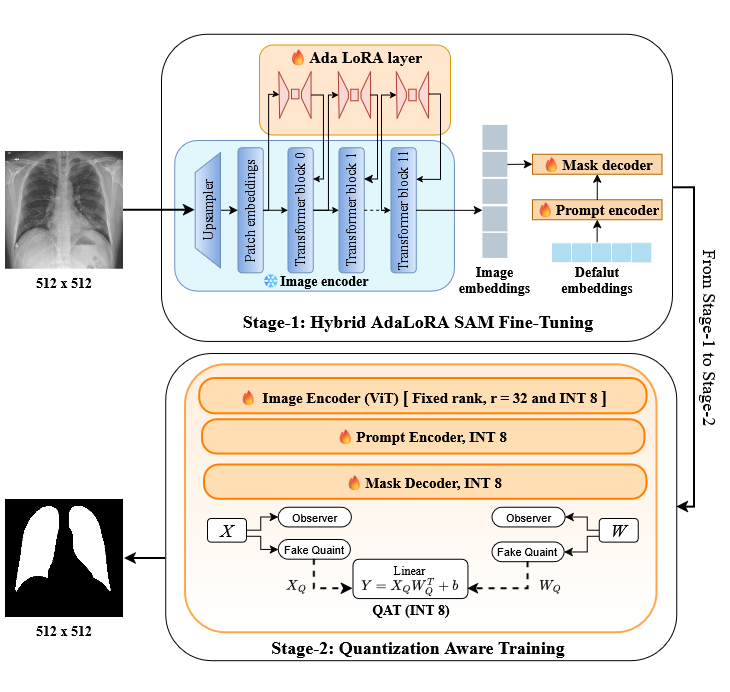}
  \vspace{-1em}
  \caption{Two-stage training pipeline of the proposed AdaLoRA-QAT framework.}
  \label{fig:pipeline_model}
\end{figure}
\vspace{-1.5em}
%-----------------------------------------------------------

\subsection{Model Overview}
\label{sec:model}

Given a pretrained weight matrix $\mathbf{W} \in \mathbb{R}^{d\times d}$, AdaLoRA models a low-rank residual $\Delta \mathbf{W}$ following the LoRA paradigm ~\cite{hu2022lora}, but dynamically allocates rank capacity across layers based on task sensitivity:
\begin{equation}
\Delta \mathbf{W}_{\text{AdaLoRA}} = \mathbf{P}\,\mathrm{diag}(\mathbf{\Lambda}\odot\mathbf{m})\,\mathbf{Q},
\end{equation}
where $\mathbf{P},\mathbf{Q}\in\mathbb{R}^{d\times r_{\max}}$ are orthogonal bases, $\mathbf{\Lambda}$ are learnable singular values, and $\mathbf{m}\in\{0,1\}^{r_{\max}}$ selects informative components. Rank importance is computed as $I_i=|\lambda_i|\cdot|\partial\mathcal{L}/\partial\lambda_i|$, retaining the top-$r_{\text{target}}$ components. Orthogonal regularization stabilizes subspace learning and prevents rank collapse.

\subsubsection{Stage 1: Full-Precision AdaLoRA Training}
AdaLoRA parameters $(\mathbf{P},\mathbf{Q},\mathbf{\Lambda},\mathbf{m})$ are optimized in FP32 using a hybrid loss:
\begin{equation}
\mathcal{L}_{\text{stage1}}=\mathcal{L}_{\text{BCE}}+\mathcal{L}_{\text{Dice}}+\lambda_{\text{ortho}}\mathcal{L}_{\text{ortho}},
\end{equation}
yielding an adaptively pruned, task-specific low-rank subspace that initializes quantization-aware fine-tuning.

\subsubsection{Stage 2: Quantization-Aware Fine-Tuning}
In Stage~2, selective mixed-precision quantization is applied: encoder linear layers (excluding attention QKV projections), decoder, and prompt encoder are quantized to INT8 using symmetric per-tensor quantization, while attention QKV projections and AdaLoRA parameters remain in FP32 to preserve orthogonality in SVD-parameterized layers. Rank masks $\mathbf{m}$ are frozen, and only $\mathbf{\Lambda}$ is updated to compensate for quantization-induced shifts, yielding:
\begin{equation}
\begin{aligned}
\mathbf{W}_{\text{qkv}},\{\mathbf{P},\mathbf{Q},\mathbf{\Lambda}\} &\in \text{FP32},\\
\mathbf{W}_{\text{enc}}^{\setminus\text{qkv}},\mathbf{W}_{\text{dec}},\mathbf{W}_{\text{prompt}} &\in \text{INT8}.
\end{aligned}
\label{eq:mixed_precision}
\end{equation}

The QAT objective is defined as,
\begin{equation}
\mathcal{L}_{\text{QAT}}=\mathcal{L}_{\text{BCE}}+\mathcal{L}_{\text{Dice}},
\quad
\text{s.t. } \partial\mathcal{L}/\partial\mathbf{m}=0,
\label{eq:qat_loss}
\end{equation}
ensuring preservation of learned subspace topology while adapting to quantization noise.
%--------------------------------------------------
\subsection{Training Strategy}
\label{sec:training}
All experiments were conducted on NVIDIA RTX~A6000 GPUs (48~GB) using an 80:10:10 split. In Stage~1, AdaLoRA adapts the vision encoder with rank reduced from 48 to 32 via importance-based pruning at epochs 3, 7, and 12, while the mask decoder is fully fine-tuned using differential learning rates ($5{\times}10^{-5}$ encoder, $2{\times}10^{-5}$ decoder), batch size 16, and orthogonality regularization ($\lambda_{\text{ortho}}{=}0.003$). In Stage~2, INT8 quantization is applied, rank masks are frozen, and only singular values $\mathbf{\Lambda}$ are fine-tuned at $1{\times}10^{-6}$, achieving substantial compression while preserving Dice performance.

%--------------------------------------------------
\section{Results and Discussion}
\label{sec:results}
\begin{table}[!t]
\centering
\caption{Quantitative comparison of segmentation performance across baseline models and SAM Ada-LoRA + Full QAT model.}
\label{tab:segmentation_results}
\resizebox{\columnwidth}{!}{
\begin{tabular}{lccc}
\toprule
\textbf{Model} & \textbf{DSC} & \textbf{IOU} & \textbf{NSD} \\
\midrule
nnU-Net & 0.9485 $\pm$ 0.119 & 0.9184 $\pm$ 0.148 & 0.8118 $\pm$ 0.233 \\
DeepLabv3+ & 0.9536 $\pm$ 0.117 & 0.9269 $\pm$ 0.142 & 0.8405 $\pm$ 0.213 \\
SegFormer & 0.9228 $\pm$ 0.174 & 0.8890 $\pm$ 0.202 & 0.7622 $\pm$ 0.237 \\
SAM Decoder & 0.9555 $\pm$ 0.0664 & 0.9210 $\pm$ 0.0998 & 0.9279 $\pm$ 0.0956 \\
SAM Ada-LoRA + D-QAT $^{\ast}$ & \textbf{0.9563 $\pm$ 0.0651} & \textbf{0.9224 $\pm$ 0.0984} & \textbf{0.9293 $\pm$ 0.0947} \\
SAM Ada-LoRA + Full QAT $^{\dagger}$ & \textbf{0.9559 $\pm$ 0.0647} & \textbf{0.9216 $\pm$ 0.0983} & \textbf{0.9286 $\pm$ 0.0938} \\
\bottomrule
\end{tabular}}

\raggedright
$^{\ast}$ {\footnotesize SAM Ada-LoRA + Decoder-only QAT Model;}
$^{\dagger}$ {\footnotesize Proposed SAM Ada-LoRA + Full QAT method; Wilcoxon signed-rank test shows no significant difference from SAM Decoder baseline (p $>$ 0.05 for all metrics), confirming performance preservation via full quantization.}
\end{table}
%--------------------------------------------------
\begin{table}[hbt]
\centering
\begin{threeparttable}
\caption{Parameter efficiency and segmentation performance comparison.}
\label{tab:model_comparison}
\setlength{\tabcolsep}{4.5pt}
\renewcommand{\arraystretch}{0.7}
\scriptsize
\begin{tabular}{lcccc}
\toprule
\textbf{Model} & \textbf{Total} & \textbf{Trainable} & \textbf{Reduction} & \textbf{DSC} \\
               & \textbf{(M)}   & \textbf{(M)}       &                    & \textbf{(\%)} \\
\midrule
nnU-Net        & 30.0 & 30.0 & -- & 94.85 \\
DeepLabV3+     & 41.0 & 41.0 & -- & 95.36 \\
SegFormer      & 84.6 & 84.6 & -- & 92.28 \\
\midrule
SAM Decoder FT & 89.7 & 3.8  & 23.6× & 95.55 \\
\midrule
\textbf{Ours (SAM Ada-LoRA)}     & 89.7 & 5.4 & 16.6× & 95.60 \\
\textbf{Ours (SAM Ada-LoRA + D-QAT)} & 89.7 & 5.4 & 16.6× & 95.63 \\
\textbf{Ours (SAM Ada-LoRA + Full QAT)} & 89.7 & 5.4 & 16.6× & 95.59 \\
\bottomrule
\end{tabular}
\begin{tablenotes}
\footnotesize
\item D-QAT = Decoder-only QAT\, ; \quad FT = Fine tuning
\end{tablenotes}
\end{threeparttable}
%\vspace{-3mm}
\end{table}

%------------------------------
\subsection{Quantitative and Qualitative Evaluation}
Table~\ref{tab:segmentation_results} and Table~\ref{tab:model_comparison} summarize the quantitative results and parameter efficiency comparison. The proposed AdaLoRA-QAT achieves Dice score (95.59\%) while requiring only 5.4M trainable parameters with 16.6$\times$ reduction and  2.24$\times$ compression compared to base-SAM fine-tuning.

Following the MedSAM evaluation protocol~\cite{ma2024segment}, statistical significance was assessed using the Wilcoxon signed-rank test~\cite{wilcoxon1963critical}. Results indicate no statistically significant difference between the proposed method and the SAM Decoder baseline across all metrics ($p>0.05$), confirming that full INT8 quantization preserves segmentation accuracy.
%-------------------------------------------------------
%--------------

%-----------------------------------------
% SSIM
\begin{figure}[htb]
  \centering
  \includegraphics[width=8.5cm]{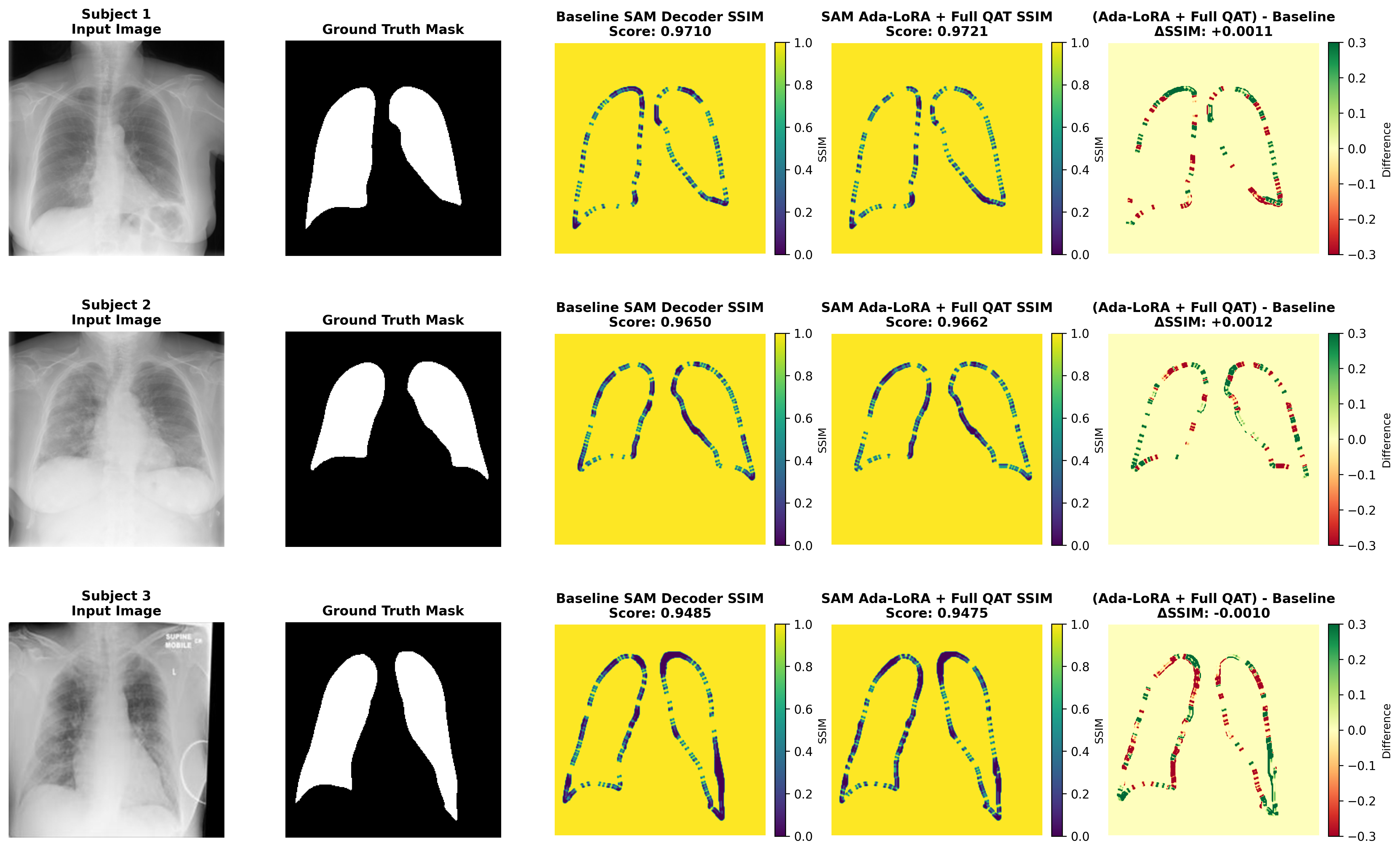}
  \caption{Structural Similarity Index (SSIM) heatmap comparison of lung segmentation.
From left: input CXR, ground truth, baseline SAM SSIM map, proposed AdaLoRA+ Full QAT SSIM map, and $\Delta$SSIM (QAT -- Baseline).
Bright regions indicate higher structural agreement.
Green in the difference map denotes localized improvements, while red marks degradations.}
\label{fig:ssim_analysis}
\end{figure}
%----------------------------------

Figure~\ref{fig:ssim_analysis} presents SSIM heatmap comparisons between baseline SAM and AdaLoRA+Full QAT. The proposed method exhibits stronger structural agreement along lung boundaries and vascular regions. The $\Delta$SSIM map shows localized improvements in low-contrast regions, with minor degradations primarily associated with severe motion artifacts or extreme pathologies, demonstrating preserved anatomical fidelity under INT8 compression.

%---------------------------------------------

\subsection{Quantization Error and Statistical Validation}
\label{sec:quant_error}
\begin{figure}[htb]
  \centering
  \includegraphics[width=8.5cm]{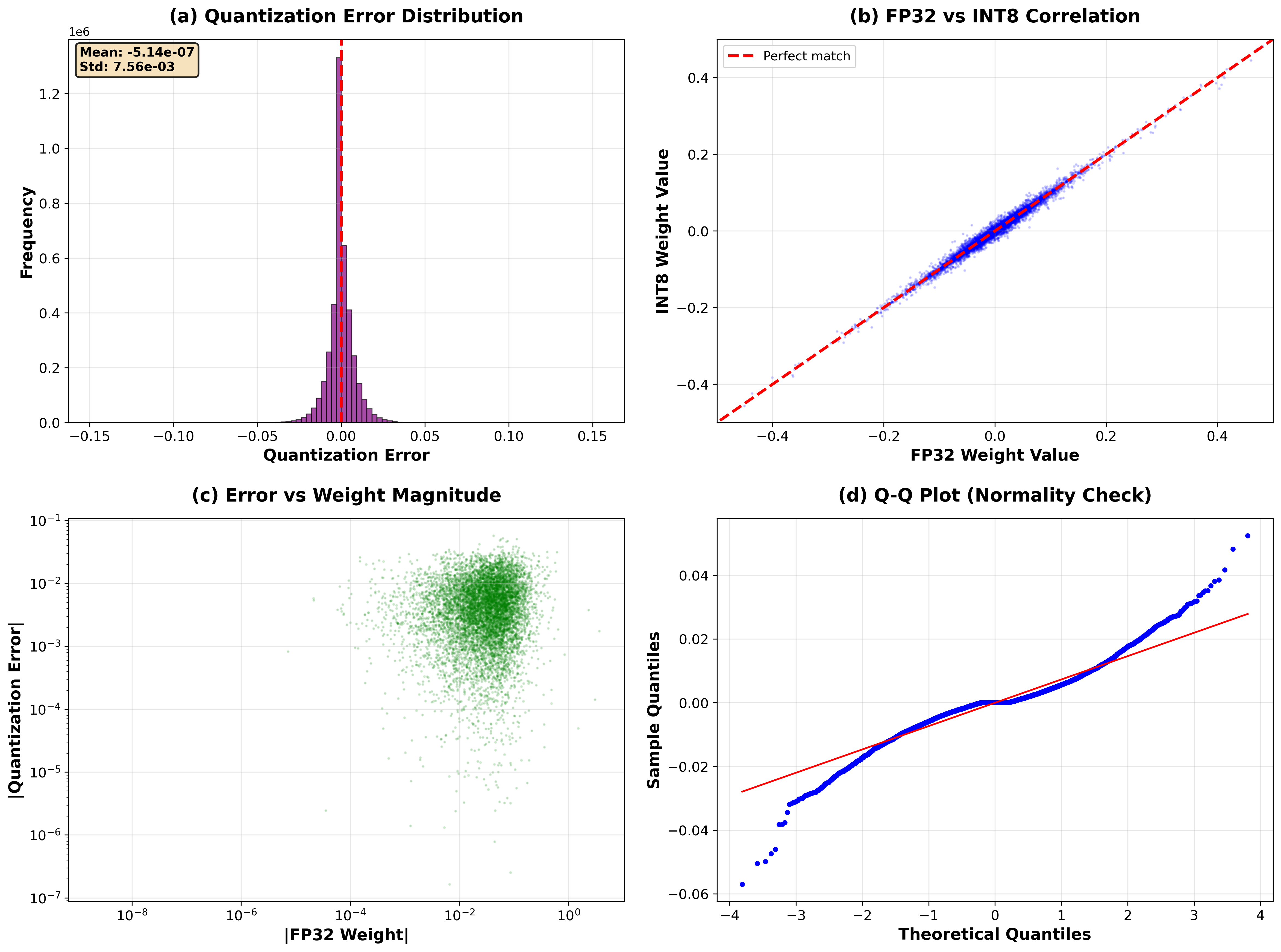}
  \caption{Quantization error analysis of AdaLoRA-QAT: (a) zero-mean Gaussian noise distribution, (b) FP32-–INT8 correlation, (c) stable error across weight amplitudes, and (d) Q–Q validation of normality.}
  \label{fig:quantization_error_analysis}
\end{figure}
Figure~\ref{fig:quantization_error_analysis} shows that FP32--INT8 quantization noise in AdaLoRA -QAT follows an approximately zero-mean Gaussian distribution ($\mu\!\approx\!2.7{\times}10^{-5}$, $\sigma\!\approx\!7.8{\times}10^{-3}$), indicating unbiased rounding behavior. Strong linear correlation between FP32 and INT8 weights and uniformly distributed errors across weight magnitudes confirm preserved numerical fidelity under low-bit quantization. Q--Q normality analysis further validates stable gradient propagation during training. Collectively, these results demonstrate that AdaLoRA-QAT achieves statistically robust INT8 compression without degrading segmentation performance.
%------------------------------------------------
%-----------------------------------------------------------
\begin{table}[t]
\centering
\caption{Component-wise ablation of the SAM-base model with AdaLoRA (rank = 32).}
\label{tab:component_ablation}
\setlength{\tabcolsep}{3.0pt}
\renewcommand{\arraystretch}{1.1}
\scriptsize
\begin{tabular}{lcccc}
\toprule
\textbf{Method} & \textbf{Encoder} & \textbf{Decoder} & \textbf{Train (\%)} & \textbf{Dice (\%)} \\
\midrule
Decoder-only FT & Frozen & Full  & 4.2 & 95.55 \\
Enc.-only LoRA (r=8) & LoRA & Frozen & 1.9 & 70.86 \\
Hybrid (Stage~1) & AdaLoRA & Full & 6.1 & 95.60 \\
Hybrid + D-QAT (Stage~2) & AdaLoRA & INT8 & 6.1 & 95.63 \\
Hybrid + Full QAT  & AdaLoRA and INT8 & INT8 & 6.1 & 95.59 \\
\bottomrule
\end{tabular}
\begin{tablenotes}
\footnotesize
\item Train (\%) shows the fraction of trainable parameters; The ``Hybrid'' configuration updates a small subset of parameters while preserving accuracy; D-QAT \, = \, Decoder-only QAT.
\end{tablenotes}
\end{table}
%------------------------------------------------

\subsection{Ablation Study}

Table~\ref{tab:component_ablation} presents component-wise ablations. Decoder-only fine-tuning achieves 95.55\% Dice with 4.2\% trainable parameters but freezes the encoder. Encoder-only LoRA ($r{=}8$) severely degrades performance (70.86\% Dice), highlighting the necessity of decoder adaptation for spatial precision. The proposed Stage~1 hybrid jointly adapts encoder and decoder, achieving 95.60\% Dice with 6.1\% trainable parameters. Extending to full QAT preserves accuracy (95.59\%, $\Delta{=}0.01\%$) while enabling 2.24$\times$ compression for efficient INT8 deployment.

Fixed-rank LoRA ($r\!\in\!\{8,16,32\}$) improves representation with increasing rank but introduces redundancy and reduced efficiency. In contrast, AdaLoRA employs SVD-based reparameterization with sensitivity-guided pruning to adaptively allocate layer-wise rank capacity, preserving essential components while maintaining compactness.
%
%-------------------------------------------------------------

%---------------------------------------------
% -------------------------------------------------------------------
\section{Conclusion}
\label{sec:conclusion}
We propose AdaLoRA-QAT, an efficient adaptation of foundation models for medical image segmentation under strict computational constraints, achieving 95.6\% Dice with 16.6$\times$ parameter reduction and 2.24$\times$ model compression while preserving anatomical fidelity. The key contribution is a mixed-precision strategy that retains attention and AdaLoRA parameters in FP32 while quantizing remaining components to INT8, preventing rank collapse in SVD-parameterized layers and enabling deployment on resource-constrained clinical hardware.

This work establishes a proof of concept for large-scale chest X-ray lung segmentation within a task-agnostic framework extensible to other medical imaging domains. While demonstrating state-of-the-art parameter efficiency, future work will explore deeper quantization, cross-modality and multi-organ validation, hardware deployment analysis, and prospective clinical evaluation, showing that foundation models can be substantially compressed without compromising diagnostic accuracy for scalable AI-assisted diagnosis in low-resource healthcare settings.
% ------------------------------------------------------------------------- 
\bibliographystyle{IEEEtran}
\bibliography{strings,refs}

@article{mittal2017lung,
  title={Lung field segmentation in chest radiographs: a historical review, current status, and expectations from deep learning},
  author={Mittal, Ajay and others},
  journal={IET Image Processing},
  volume={11},
  number={11},
  pages={937--952},
  year={2017},
  publisher={Wiley Online Library}
}

@inproceedings{chen2018encoder,
  title={Encoder-decoder with atrous separable convolution for semantic image segmentation},
  author={Chen, Liang-Chieh and others},
  booktitle={ECCV},
  pages={801--818},
  year={2018}
}

@article{isensee2018nnu,
  title={nnu-net: Self-adapting framework for u-net-based medical image segmentation},
  author={Isensee, Fabian and others},
  journal={arXiv preprint arXiv:1809.10486},
  year={2018}
}

@article{xie2021segformer,
  title={SegFormer: Simple and efficient design for semantic segmentation with transformers},
  author={Xie, Enze and others},
  journal={Advances in neural information processing systems},
  volume={34},
  pages={12077--12090},
  year={2021}
}

@inproceedings{kirillov2023segment,
  title={Segment anything},
  author={Kirillov, Alexander and others},
  booktitle={Proceedings of the IEEE/CVF international conference on computer vision},
  pages={4015--4026},
  year={2023}
}

@article{hu2022lora,
  title={Lora: Low-rank adaptation of large language models.},
  author={Hu, Edward J and others},
  journal={ICLR},
  volume={1},
  number={2},
  pages={3},
  year={2022}
}

@article{shiraishi2000development,
  title={Development of a digital image database for chest radiographs with and without a lung nodule: receiver operating characteristic analysis of radiologists' detection of pulmonary nodules},
  author={Shiraishi, junji and others},
  journal={American journal of roentgenology},
  volume={174},
  number={1},
  pages={71--74},
  year={2000},
  publisher={American Roentgen Ray Society}
}

@inproceedings{degerli2022osegnet,
  title={Osegnet: Operational segmentation network for covid-19 detection using chest x-ray images},
  author={Degerli, Aysen and others},
  booktitle={ICIP},
  pages={2306--2310},
  year={2022},
  organization={IEEE}
}

@article{rahman2021exploring,
  title={Exploring the effect of image enhancement techniques on COVID-19 detection using chest X-ray images},
  author={Rahman, Tawsifur and others},
  journal={Computers in biology and medicine},
  volume={132},
  pages={104319},
  year={2021},
  publisher={Elsevier}
}

@article{chowdhury2020can,
  title={Can AI help in screening viral and COVID-19 pneumonia?},
  author={Chowdhury, Muhammad EH and others},
  journal={Ieee Access},
  volume={8},
  pages={132665--132676},
  year={2020},
  publisher={IEEE}
}

@article{zhang2023adalora,
  title={Adalora: Adaptive budget allocation for parameter-efficient fine-tuning},
  author={Zhang, Qingru and others},
  journal={arXiv preprint arXiv:2303.10512},
  year={2023}
}

@article{tahir2021covid,
  title={COVID-19 infection localization and severity grading from chest X-ray images},
  author={Tahir, Anas M and others},
  journal={Computers in biology and medicine},
  volume={139},
  pages={105002},
  year={2021},
  publisher={Elsevier}
}

@misc{siim-acr-pneumothorax,
  author       = {Anna Zawacki and others},
  title        = {SIIM-ACR Pneumothorax Segmentation},
  howpublished = {\emph{Kaggle}, [Online]. Available: \url{https://www.kaggle.com/competitions/siim-acr-pneumothorax-segmentation}},
  year         = {2019}
}

@article{ma2024segment,
  title={Segment anything in medical images},
  author={Ma, Jun and others},
  journal={Nature Communications},
  volume={15},
  number={1},
  pages={654},
  year={2024},
  publisher={Nature Publishing Group UK London}
}

@book{wilcoxon1963critical,
  title={Critical values and probability levels for the Wilcoxon rank sum test and the Wilcoxon signed rank test},
  author={Wilcoxon, Frank and others},
  volume={1},
  year={1963},
  publisher={American Cyanamid Pearl River, NY}
}

\section{Acknowledgement}
The work was supported by IHub-Data, International Institute of Information Technology Hyderabad. Tapabrata Chakraborti is supported by the Turing-Roche Strategic Partnership and the UCL NIHR Biomedical Research Center.

\section{COMPLIANCE WITH ETHICAL STANDARDS}
\label{ethical}
The data used are all from public benchmark datasets for which ethical approvals were already pre-existing from the original studies that collected them. The present work is a computational simulation study on that anonymised open access data for which no further ethical approval was required. 

\section{Conflicts of Interest}
\label{conflict}
The authors have no conflicts of interest.

\end{document}